\documentclass[aps,prl,twocolumn,showpacs,superscriptaddress]{revtex4}
\newcommand{\PRE}[1]{}       % Use if journal style
\usepackage{epsfig}
\newcommand{\be}{\begin{equation}}
\newcommand{\ee}{\end{equation}}
\newcommand{\beas}{\begin{eqnarray*}}
\newcommand{\eeas}{\end{eqnarray*}}
\newcommand{\bea}{\begin{eqnarray}}
\newcommand{\eea}{\end{eqnarray}}

\def\av#1{\langle #1\rangle}
\def\neg{N_{e \gamma}}
\def\bn{N-\neg}

\newcommand{\bwide}{\begin{widetext}}
\newcommand{\ewide}{\end{widetext}}
\newcommand{\postscript}[2]{\setlength{\epsfxsize}{#2\hsize}
   \centerline{\epsfbox{#1}}}

\newcommand{\mplanck}{M_{\text{Pl}}}

\newcommand{\md}{M_D}
\newcommand{\mbh}{M_{\text{BH}}}
\newcommand{\Tbh}{T_{\text{BH}}}
\newcommand{\mbhmin}{M_{\text{BH}}^{\text{min}}}

\newcommand{\xmin}{x_{\text{min}}}
\newcommand{\ifb}{\text{fb}^{-1}}
\newcommand{\iab}{\text{ab}^{-1}}

\newcommand{\gev}{\text{GeV}}
\newcommand{\tev}{\text{TeV}}

\newcommand{\eqref}[1]{Eq.~(\ref{#1})}

\begin{document}

\preprint{
\hfil
\begin{minipage}[t]{3in}
\begin{flushright}
\vspace*{.4in}
NUB--3243--TH--03\\
UCI--TR--2003--30\\
UK/03--15\\
hep-ph/0311365
\end{flushright}
\end{minipage}
}

\title{
\PRE{\vspace*{1.5in}}
Inelastic Black Hole Production and Large Extra Dimensions
\PRE{\vspace*{0.3in}}
}

\author{Luis A.~Anchordoqui}
\affiliation{Department of Physics,
Northeastern University, Boston, MA 02115
\PRE{\vspace*{.1in}}
}

\author{Jonathan L.~Feng}
\affiliation{Department of Physics and Astronomy,
University of California, Irvine, CA 92697
\PRE{\vspace*{.1in}}
}

\author{Haim Goldberg}
\affiliation{Department of Physics,
Northeastern University, Boston, MA 02115
\PRE{\vspace*{.1in}}
}

\author{Alfred D.~Shapere}%
\affiliation{Department of Physics,
University of Kentucky, Lexington, KY 40502
\PRE{\vspace*{.5in}}
}

\date{November 2003}

\begin{abstract}
\PRE{\vspace*{.1in}} Black hole production in elementary particle
collisions is among the most promising probes of large extra spacetime
dimensions.  Studies of black holes at particle colliders have assumed
that all of the incoming energy is captured in the resulting black
hole.  We incorporate the inelasticity inherent in such processes and
determine the prospects for discovering black holes in colliders and
cosmic ray experiments, employing a dynamical model of Hawking
evolution.  At the Large Hadron Collider, inelasticity reduces rates
by factors of $10^3$ to $10^6$ in the accessible parameter space,
moderating, but not eliminating, hopes for black hole discovery.  At
the Pierre Auger Observatory, rates are suppressed by a factor of 10.
We evaluate the impact of cosmic ray observations on collider
prospects.
\end{abstract}

%04.70.-s   Physics of black holes
%04.50.+h   Gravity in more than four dimensions
%13.85.Qk   Hadron-induced high- and super-high-energy interactions:
%             Inclusive production with identified leptons, photons,
%             or other nonhadronic particles
%96.40.Tv   Neutrino and muon cosmic rays

\maketitle

When the Large Hadron Collider (LHC) experiences ``first light'' later
this decade, some of this ``light'' may be the Hawking radiation of
microscopic black holes
(BHs)~\cite{Banks:1999gd,Giddings:2001bu,Dimopoulos:2001hw}. Likewise,
ultrahigh energy cosmic rays that are continuously bombarding the
Earth may be producing several BHs per minute in the upper
atmosphere~\cite{Feng:2001ib}, as well as in the Antarctic ice
cap~\cite{Kowalski:2002gb}.

The production of microscopic BHs is possible if large extra spacetime
dimensions exist.  In the simplest scenarios, spacetime is a direct
product of the apparent 4-dimensional world, where standard model (SM)
fields and gravity propagate, and a flat spatial $n$-dimensional torus
(of common linear size $2\pi r_c$), where only gravity
propagates~\cite{Arkani-Hamed:1998rs}. In such a spacetime, gravity is
modified at distances below $r_c$ and becomes strong at the energy
scale $\md \equiv [\mplanck^2 / (8 \pi r_c^n)]^{1/(2+n)}$, where
$\mplanck \sim 10^{19}~\gev$ is the 4-dimensional Planck
scale. Gravitational collapse may therefore be triggered in collisions
of elementary particles with center-of-mass energy somewhat above
$\md$~\cite{Eardley:2002re} at small impact parameters.  Astrophysical
constraints require $\md \gg 10~\tev$ for $n= 2,3$ and $\md \agt
4~\tev$ for $n=4$~\cite{Hagiwara:fs}.  For $n \geq 5$, however, $\md$
may be as low as a TeV~\cite{Abbott:2000zb,Anchordoqui:2003jr}, and so
BH production is possible in $pp$ collisions at the LHC.

Up to now,  studies~\cite{Dimopoulos:2001hw,Ringwald:2001vk} of
BH production at particle colliders have
taken the mass trapped inside the BH's apparent horizon, $\mbh$, to be
identical to the incoming parton energy $\sqrt{\hat{s}}$.  This is a
poor approximation: even in head-on collisions at impact parameter
$b=0$, a significant fraction of the incoming energy may escape in
gravitational waves.  Recently, Yoshino and Nambu have quantified the
inelasticity $y \equiv M_{\rm BH}/\sqrt{\hat s}$ as a function of $n$
and $b$~\cite{Yoshino:2002tx}.  Although Yoshino and Nambu have only
determined lower bounds on $y$, their results may still be taken as
reasonable estimates of the effects of inelasticity.  In this work we
include inelasticity and determine its effect on BH discovery
prospects at the LHC.

Including inelasticity, the BH cross section at a hadron collider with
center-of-mass energy $\sqrt{s}$ is
\begin{eqnarray}
\lefteqn{\sigma^{pp} (s, x_{\text{min}}, n, \md) \equiv
\int_0^1 \! \! 2 z \, dz
\int_{\frac{(x_{\text{min}} \md)^2}{y^2 s}}^1 \! \! du \int_u^1 \! \!
\frac{dv}{v} } \ \nonumber \\
&& \times F(n) \, \pi r_s^2(u s, n, \md) \sum_{ij} f_i(v,Q) f_j(u/v,Q)
\, ,
\label{sigma}
\end{eqnarray}
where $z = b/b_{\text{max}}$, $F(n)$ is the form factor of
Ref.~\cite{Yoshino:2002tx},
\begin{equation}
r_s (us, n, \md) = k(n) \md^{-1}
\left[ \sqrt{us}/ \md \right]^{1/(1+n)} \, ,
\end{equation}
where
\begin{equation}
k(n) \equiv \left[
2^n \sqrt{\pi}^{n-3} \frac{\Gamma[(3+n)/2]}{2+n} \right] ^{1/(1+n)}
\, ,
\end{equation}
is the Schwarzschild radius of the apparent horizon~\cite{Myers:un},
$i,j$ label parton species, $f_i, f_j$ are parton distribution
functions~\cite{Pumplin:2002vw} with momentum transfer $Q \simeq
r_s^{-1}$~\cite{Emparan:2001kf}, and $x_{\text{min}}=\mbhmin/\md$,
where $\mbh^{\rm min}$ is the smallest BH mass for which we trust the
semi-classical calculation.

\begin{figure}
\postscript{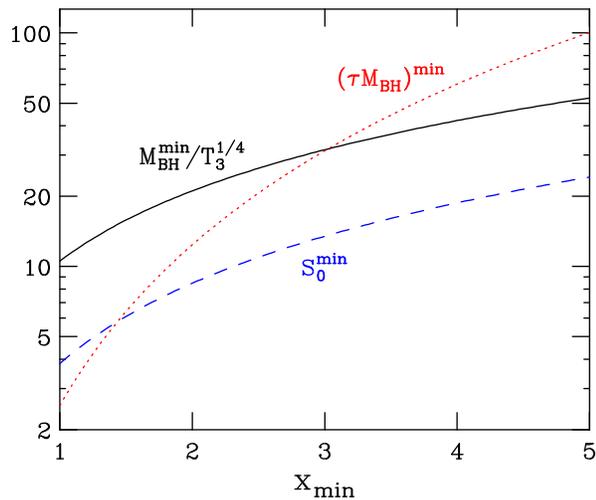}{0.90}
\caption{Quantitative measures of the validity of the semi-classical
  analysis of BH production for $n=6$. (See text.)}
\label{fig:xminplot}
\end{figure}

The parameter $\xmin$ plays in important role in interpreting the
results derived below.  Validity of the semi-classical calculation
requires satisfaction of at least three
criteria~\cite{Giddings:2001bu}.  First, $S_0$, the initial entropy of
the produced BH, should be large enough to ensure a well-defined
thermodynamic description~\cite{Preskill:1991tb}. Second, the BH's
lifetime $\tau$ should be large compared to its inverse mass so that
the black hole behaves like a well-defined resonance.  Third, the BH's
mass must be large compared to the scale of the 3-brane tension $T_3$
so that the brane does not significantly perturb the BH metric.
Quantitative measures of these three criteria are given in
Fig.~\ref{fig:xminplot}, assuming $T_3 = \sqrt{8\pi} /(2\pi)^6 \md^4$
for 6 toroidally-compactified dimensions~\cite{Polchinski:1996na}.  We
find it reasonable to conclude that all three criteria are satisfied
for $\xmin \approx 3$, but not necessarily for lower $\xmin$. In
string theory, as the BH mass decreases toward $\md$, there is a
continuous transition, at least in energy and string coupling, to
string ball production~\cite{Dimopoulos:2001qe}.  In many string
models, the string ball cross section lies well above the
semi-classical BH cross section, perhaps justifying extrapolation to
$\xmin \approx 1$~\cite{Anchordoqui:2003jr}.

\begin{figure}
\postscript{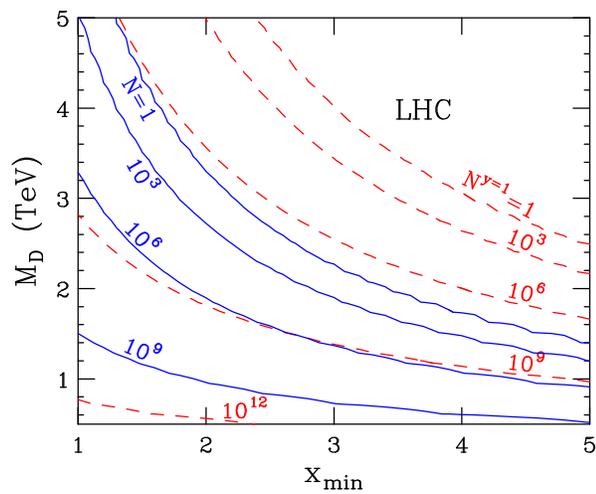}{0.90}
\caption{The number of BHs produced at the LHC with
inelasticity included ($N$, solid) and neglected ($N^{y=1}$, dashed)
for integrated luminosity $1~\iab$ and $n=6$ extra dimensions.}
\label{fig:LHCcombo}
\end{figure}

The number of BHs produced at the LHC with and without inelasticity is
given in Fig.~\ref{fig:LHCcombo} in the $(\xmin, \md)$ plane, assuming
a cumulative integrated luminosity of $1~\iab$ over the life of the
collider.  Inelasticity suppresses event rates by factors of $10^3$ to
$10^6$ in the region of parameter space where more than 1 inelastic BH
event is expected.  The effect is large because the LHC is
energy-limited for BH production, and inelasticity effectively
suppresses the available energy to below BH production threshold in
much of parameter space.

\begin{figure}
\postscript{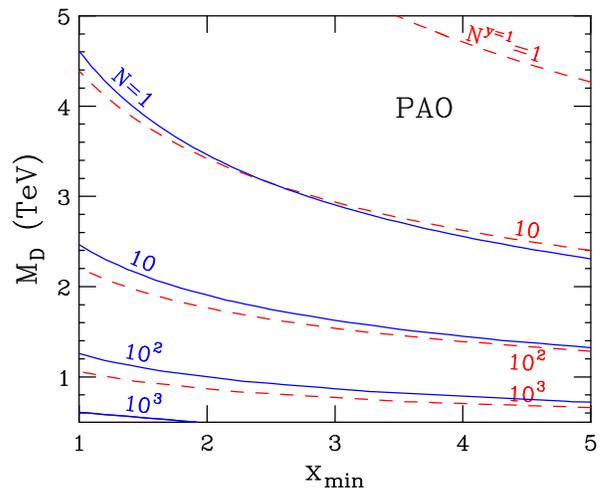}{0.90}
\caption{The number of BHs produced at the PAO with inelasticity
included ($N$, solid) and neglected ($N^{y=1}$, dashed) for 10 years
of running and $n=6$ extra dimensions. The hadronic aperture of the
PAO is given in Ref.~\cite{Capelle:1998zz}. }
\label{fig:Augercombo}
\end{figure}

Inelasticity also affects event rates for BH-mediated showers at
cosmic ray facilities. These showers are initiated by very high energy
neutrinos in the atmosphere, and the observable event rate is a
function of BH cross section, exposure, and the incoming neutrino
flux~\cite{Anchordoqui:2003jr}. In this case, inelasticity affects not
only the BH production cross section, but also the exposure, which is
a function of shower energy. We have presented the effect of
inelasticity on existing cosmic ray data
elsewhere~\cite{Anchordoqui:2003jr}. Following that analysis, we show
event rates for a future cosmic ray experiment, the Pierre Auger
Observatory (PAO), in Fig.~\ref{fig:Augercombo}, assuming a cosmogenic
flux, again with and without inelasticity.  Cosmic ray experiments are
flux-, not energy-, limited.  Consequently, the effect of inelasticity
is much less severe than at the LHC, typically reducing event rates by
an order of magnitude.

We turn now to a detailed evaluation of BH discovery prospects at the
LHC. Following Dimopoulos and Landsberg~\cite{Dimopoulos:2001hw}, we
consider the signal of events with total multiplicity $N\ge 4$ and at
least one $e^{\pm}/\gamma$ with energy $> 100~\gev$.  To implement
these cuts, we first determine average multiplicities $\av{N}$ for the
various particle species, incorporating evolution effects during
Hawking radiation.

The average total emission rate for particle species $i$ is
\begin{equation}
\frac{d \av{N} }{dt}
= \frac{1}{2\pi} \left( \sum c_i\, g_i\, \Gamma_i \right)
\zeta(3)\, \Gamma(3)\, r^2\, T^3 \, ,
\label{N}
\end{equation}
where $c_i$ is the number of internal degrees of freedom of particle
species $i$, $g_i= 1\, (3/4)$ for bosons (fermions),
\begin{equation}
\Gamma_i = \frac{1}{4\pi r^2}
\int \frac{\sigma_i(\omega)\, \omega^2\, d\omega}
{e^{\omega/T}\pm 1} \left[ \int \frac{\omega^2\, d\omega}
{e^{\omega/T}\pm 1}\right]^{-1} ,
\label{gamma}
\end{equation}
where $\sigma_i$ is the greybody absorption area due to the
backscattering of part of the outgoing radiation of frequency $\omega$
into the BH~\cite{Kanti:2002nr}, and $r$ and $T$ are the instantaneous
Schwarzschild radius and Hawking temperature, which are related by
\begin{equation}
r = \frac{1+n}{4 \pi T} = \frac{k(n)}{\md}
\left(\frac{M}{\md}\right)^{\frac{1}{1+n}} \ .
\end{equation}
The rate of change of the BH mass is
\begin{equation}
 \frac{dM}{dt} = -\frac{1}{2\pi} \left(\sum c_i\, f_i\, \Phi_i\right)
 \zeta(4)\, \Gamma(4)\, r^2\, T^4\, ,
\label{M}
\end{equation}
where $f_i = 1\, (7/8)$ for bosons (fermions) and
\begin{equation}
\Phi_i = \frac{1}{4\pi r^2} \int \frac{\sigma_i(\omega)\,
\omega^3\, d\omega}
{e^{\omega/T}\pm 1} \left[ \int \frac{\omega^3\, d\omega}
{e^{\omega/T}\pm 1} \right]^{-1} .
\label{phi}
\end{equation}
Without the absorption correction, $\Gamma_i = \Phi_i = 1$. With it,
these greybody parameters are given in Table~\ref{t}.

\begin{table}
\caption{Degrees of freedom of particle species and greybody
parameters as defined in Eqs.~(\ref{gamma}) and (\ref{phi}).}
\begin{tabular}{|c|c|c|c|}
\hline
\hspace{0.4cm} particle's spin \hspace{0.4cm} &
$\hspace{0.7cm} c_i \hspace{0.7cm}$ &
$ \hspace{0.7cm} \Gamma_i \hspace{0.7cm} $ &
$\hspace{0.7cm} \Phi_i \hspace{0.7cm} $   \\
\hline
{\small 0} & 1 & 0.80  & 0.80 \\
$\frac{1}{2}$ & 90 & 0.66 & 0.62 \\
{\small 1} & 27 & 0.60 & 0.67  \\
\hline
\end{tabular}
\label{t}
\end{table}

Dividing \eqref{N} by \eqref{M} and integrating, we obtain a compact
expression for the average multiplicity~\cite{Cavaglia:2003hg}
\begin{equation}
\av{N} = \frac{4 \pi \, \rho\, k(n)}{2+n}
\left[ \frac{\mbh}{\md} \right]^{\frac{2+n}{1+n}}  = \rho\ S_0 \, ,
\label{multiplicity}
\end{equation}
where
\begin{equation}
\rho=\frac{\sum c_i\, g_i\, \Gamma_i}{\sum c_i\, f_i\,
\Phi_i} \frac{\zeta(3)\, \Gamma(3)} {\zeta(4)\, \Gamma(4)} \, ,
\end{equation}
and
\begin{equation}
S_0 = \left(\frac{1+n}{2+n}\right) \frac{\mbh}{\Tbh}
\end{equation}
is the initial value of the entropy in terms of the initial BH mass
and Hawking temperature $\Tbh$.  The average multiplicity for any
subset of states $\{s\}$ is $\langle N_{\{s\}}\rangle = B_{\{s\}}
\av{N}$, where the branching fraction is
\begin{equation}
B_{\{s\}} = \frac{\sum_{i \in \{s\}} c_i\, g_i\, \Gamma_i}
{\sum_i c_i\, g_i\, \Gamma_i} \, .
\label{branching}
\end{equation}
For $n=6$, using the the parameters given in Table~\ref{t}, we find
$\av{N}=0.30\, M/T$ and $\av{\neg} = 0.052 \, \av{N} = 0.016 \, M/T$.
Note that $\av{N}$ is a factor of 3 smaller than the entropy $S_0,$
and somewhat smaller than
$\av{N} =0.5\, M/T$ used in previous
calculations~\cite{Dimopoulos:2001hw,Feng:2001ib}.

$\av{N}$ is the average value of a Poisson distribution.  If all
species are Poisson distributed, then the sum of particles in any
subset is also Poisson distributed, and so $N$, $\neg$, and $N-\neg$
are all Poisson distributed, where $\neg$ is the total number of
$e^{\pm}/\gamma$ per event.  The probability that a given event has
$\neg=0$ and $N \ge 4$ is therefore
\begin{eqnarray}
\lefteqn{P(\neg =0)\, P(\bn\ge 4) =  } \quad \nonumber \\
&& e^{-\av{\neg}}\, e^{-\av{\bn}} \sum_{i\ge 4}
\frac{\av{\bn} ^i}{i\, !} \, .
\end{eqnarray}
Finally, the probability of $\neg \ge 1$ and $N \ge 4$ is
\begin{widetext}
\begin{equation}
 e^{-\av{N}}  \sum_{i\ge 4} \frac{\av{N} ^i}{i\, !} -
e^{-\av{N}}  \sum_{i\ge 4} \frac{\av{\bn} ^i}{i\, !}
  =  \left(1- e^{-\av{N}} \sum_{i=0}^{3}\frac{\av{N} ^i}{i\, !}\right)
 - e^{-\av{\neg}} \left(1 - e^{-\av{\bn}}
\sum_{i=0}^3\frac{\av{\bn} ^i}{i\, !}\right)\, . \nonumber
\end{equation}
\end{widetext}

\begin{figure}
\postscript{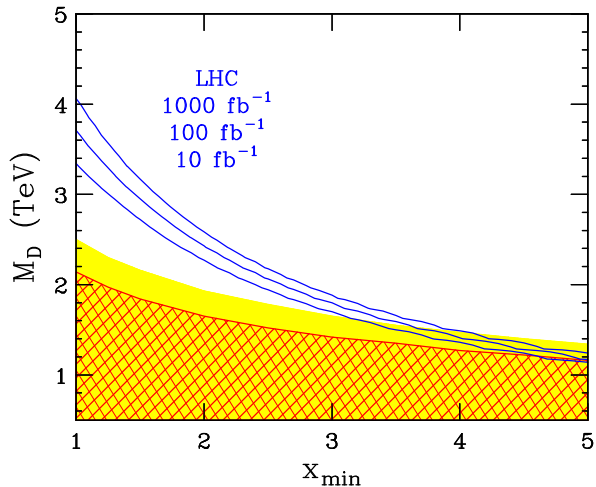}{0.90} \caption{The discovery reaches for
the LHC (solid) for 3 different integrated luminosities and $n=6$
extra dimensions. Also shown is the region of parameter space
which can be excluded at 95\% CL if no neutrino showers mediated
by BHs are observed in 5 years at the PAO. The shaded
(cross-hatched) region assumes 2 SM neutrino + 0 (10) hadronic
background events. } \label{fig:reach}
\end{figure}

The SM background masking such events is dominated by $Z(e^+e^-)$ +
jets and $\gamma$ + jets. We adopt the background rates estimated
using PYTHIA in Ref.~\cite{Dimopoulos:2001hw}, and require a 5$\sigma$
excess for discovery~\cite{Anchordoqui:2002cp}.  The resulting reach
is shown in Fig.~\ref{fig:reach}.

Since the PAO will begin operation before the LHC, it is of interest
to see what cosmic ray observations might imply for collider
prospects. In Fig.~\ref{fig:reach} we have superimposed the region of
parameter space excluded if no BH events above background are found at
the PAO.  Assuming 1 ab$^{-1}$ luminosity for the LHC and a background
of up to 2 SM neutrinos and 10 hadronic events at the PAO, we find
that the PAO can limit the discovery reach of LHC to a triangular
region in the $(\xmin, \md)$ plane, ranging from 2.2 to 4.0 TeV for
$\xmin=1$ and only from 1.4 to 1.8 TeV at the favored value
$\xmin=3$.   At
$\xmin=3$ the $\md$-sensitivity of the PAO is reduced with respect to
our previous estimate~\cite{Anchordoqui:2001cg} by a factor of 1.6 as
a result of inelasticity.

To summarize, we have analyzed the impact on event rates of
inelasticity in BH production.  The effects of inelasticity are
considerable, suppressing event rates at the LHC by factors of $10^3$
to $10^6$ in the semi-classical regime. Our dynamical treatment of
Hawking evolution also reduces event multiplicities compared to
previous estimates.  In spite of the enormous suppression, BH
discovery at the LHC is still possible. We recall also that the
trapped mass estimate in Ref.~\cite{Yoshino:2002tx} is a {\em lower
bound} on the BH mass; a change in the radiation profile, especially
at large impact parameters, could considerably raise event rates.
Finally, we have also shown that non-observation of an excess of
deeply-penetrating showers at the PAO would significantly restrict the
parameter space for BH discovery at the LHC.

\acknowledgments{The work of LAA is supported in part by
%U.S. National Science Foundation (NSF) grant No.\ PHY--0140407.  The
NSF grant No.\ PHY--0140407.  The
work of JLF is supported in part by NSF CAREER grant No.\
PHY--0239817.  The work of HG is supported in part by NSF grant No.\
PHY--0244507.  The work of ADS is supported in part
%by Department of Energy Grant No.\ DE--FG01--00ER45832 and NSF Grant PHY--0245214.}
by DOE Grant No.\ DE--FG01--00ER45832 and NSF Grant PHY--0245214.}

\end{document}